\documentclass[aps,prb, preprint,amsmath,amssymb,showpacs,superscriptaddress]{revtex4-1}

\usepackage{graphicx}
\usepackage{color}

\makeatletter
\@ifundefined{textcolor}{}
{
 \definecolor{BLACK}{gray}{0}
 \definecolor{WHITE}{gray}{1}
 \definecolor{RED}{rgb}{1,0,0}
 \definecolor{GREEN}{rgb}{0,1,0}
 \definecolor{BLUE}{rgb}{0,0,1}
 \definecolor{CYAN}{cmyk}{1,0,0,0}
 \definecolor{MAGENTA}{cmyk}{0,1,0,0}
 \definecolor{YELLOW}{cmyk}{0,0,1,0}
}

\begin{document}

\title{Discovery of Dirac Node Arcs in PtSn$_4$}

\author{Yun Wu}
\affiliation{Division of Materials Science and Engineering, Ames Laboratory, Ames, Iowa 50011, USA}
\affiliation{Department of Physics and Astronomy, Iowa State University, Ames, Iowa 50011, USA}

\author{Lin-Lin Wang}
\affiliation{Division of Materials Science and Engineering, Ames Laboratory, Ames, Iowa 50011, USA}

\author {Eundeok Mun}
\email[]{Present address: Department of Physics, Simon Fraser University, Canada}
\affiliation{Division of Materials Science and Engineering, Ames Laboratory, Ames, Iowa 50011, USA}
\affiliation{Department of Physics and Astronomy, Iowa State University, Ames, Iowa 50011, USA}

\author{D. D. Johnson}
\affiliation{Division of Materials Science and Engineering, Ames Laboratory, Ames, Iowa 50011, USA}
\affiliation{Department of Physics and Astronomy, Iowa State University, Ames, Iowa 50011, USA}
\affiliation{Department of Materials Science and Engineering, Iowa State University, Ames, Iowa 50011, USA}

\author{Daixiang Mou}

\author{Lunan Huang}
\affiliation{Division of Materials Science and Engineering, Ames Laboratory, Ames, Iowa 50011, USA}
\affiliation{Department of Physics and Astronomy, Iowa State University, Ames, Iowa 50011, USA}

\author{Yongbin Lee}
\affiliation{Division of Materials Science and Engineering, Ames Laboratory, Ames, Iowa 50011, USA}

\author{S.~L.~Bud'ko}

\author{P. C. Canfield}
\email[]{canfield@ameslab.gov}

\author{Adam Kaminski}
\email[]{kaminski@ameslab.gov}
\affiliation{Division of Materials Science and Engineering, Ames Laboratory, Ames, Iowa 50011, USA}
\affiliation{Department of Physics and Astronomy, Iowa State University, Ames, Iowa 50011, USA}

\begin{abstract}
\textbf{In topological quantum materials\cite{Hasan10RMP,Burkov11PRB, Burkov11PRL, Heikkila11JETP} the conduction and valence bands are connected at points (Dirac/Weyl semimetals) or along lines (Line Node semimetals) in the momentum space. Numbers of studies demonstrated that  several materials are indeed Dirac/Weyl semimetals\cite{Liu14Sci, Neupane14NatCom, Xu15SciDis, Xu15NatPhys, Xu15SciAdv}. However, there is still no experimental confirmation of materials with line nodes, in which the Dirac nodes form closed loops in the momentum space \cite{Burkov11PRB, Heikkila11JETP}. Here we report the discovery of a novel topological structure - Dirac node arcs - in the ultrahigh magnetoresistive material PtSn$_4$ using laser-based angle-resolved photoemission spectroscopy (ARPES) data and density functional theory (DFT) calculations. Unlike the closed loops of line nodes, the Dirac node arc structure resembles the Dirac dispersion in graphene\cite{Geim07NatMat} that is extended along one dimension in momentum space and confined by band gaps on either end. We propose that this reported Dirac node arc structure is a novel topological state that provides  a novel platform for studying the exotic properties of Dirac Fermions.}
\end{abstract}
\date{\today}
\maketitle

The discovery of non-trivial surface states in topological insulators\cite{Hasan10RMP} attracted a lot of interest and initiated quests for novel diverse topological states in condensed matter. Topological nodal states with conduction and valence bands touching at points (Dirac/Weyl semimetals) or lines (Line Node semimetals) have been proposed to exist in multilayer heterostructures\cite{Burkov11PRB, Burkov11PRL}. A possible extension of these states to three dimensional (3D) single crystals was proposed in $\beta$-cristobalite BiO$_2$ \cite{Young12PRL} and A$_3$Bi (A = Na, K, Rb)\cite{Wang12PRB}, which are thought to host bulk 3D Dirac points protected by crystal symmetry. Subsequently, Na$_3$Bi and Cd$_3$As$_2$ were experimentally demonstrated to be 3D Dirac semimetals\cite{Wang13PRB, Liu14Sci, Neupane14NatCom, Liu14NatMat, Yi14SciRep, Borisenko14PRL, Narayanan15PRL}. Subsequently, another type of massless particle  - Weyl Fermion\cite{Weyl29ZfP} - was found in states that were predicted to exist in a family of non-centrosymmetric transition metal TaAs, TaP, NbP and NbAs\cite{Huang15NatCom, Weng15PRX}. These materials were confirmed as Weyl semimetals by reports of Fermi arc states connecting the Weyl points as a unique signature\cite{Xu15SciDis, Xu15NatPhys, Xu15SciAdv}. While experimental evidence supports existence of Dirac semimetals and Weyl semimetals, clear signatures of semimetals with line nodes are yet to be discovered. Several groups proposed that line node structures may exist in graphene networks\cite{Weng15PRB}, rare earth monopnictides\cite{Zeng15arXiv}, antiperovskite Cu$_3$PdN/Cu$_3$ZnN\cite{Yu15PRL, Youngkuk15PRL}, SrIrO3\cite{Chen15PRB}, TlTaSe$_2$\cite{Bian15arXivDrum},  Ca$_3$P$_2$\cite{Xie15APLMater} and CaAg$X$($X$ = P, As)\cite{Yamakage15arXiv}, but so far no direct evidence was reported. A few ARPES studies in PbTaSe$_2$\cite{Bian15arXivTopo} and ZrSiS\cite{Schoop15arXiv} presented some evidence of the existence of Dirac-like features and ``drumhead" surface states, but further research is still needed to understand fully their significance and relation to Dirac line nodes. 

Many topological nodal semimetals such as Cd$_3$As$_2$\cite{Liang15NatMat}, NbP\cite{Shekhar15NatPhys}, WTe$_2$\cite{Ali14Nat} exhibit extremely large magnetoresistance. Prior to these discoveries, a similar effect was observed in PtSn$_4$ with magnetoresistance of $\simeq 5 \times 10^5~\%$ and no obvious saturation at 1.8K and 140 kOe\cite{Mun12PRB}. However, the band structure of this material has not been studied experimentally in detail due to very complex Fermi surface (FS) revealed by calculations\cite{Mun12PRB}. Here we demonstrate that, despite its quite complex FS in the center region of the Brillouin zone, there are also very interesting features close to boundary of the zone, i.e., the $Z$ and $X$ points, which are the signatures of a topological quantum material. Whereas most of the topological quantum materials were predicted by theory first and verified by experiment later, we present an opposite approach: we focused on ultrahigh non-saturating magnetoresistance and based on this we searched for topological states in PtSn$_4$ by using ultrahigh resolution ARPES and followed with band-structure calculations.

\begin{figure*}[tb]
	\includegraphics[scale=0.5]{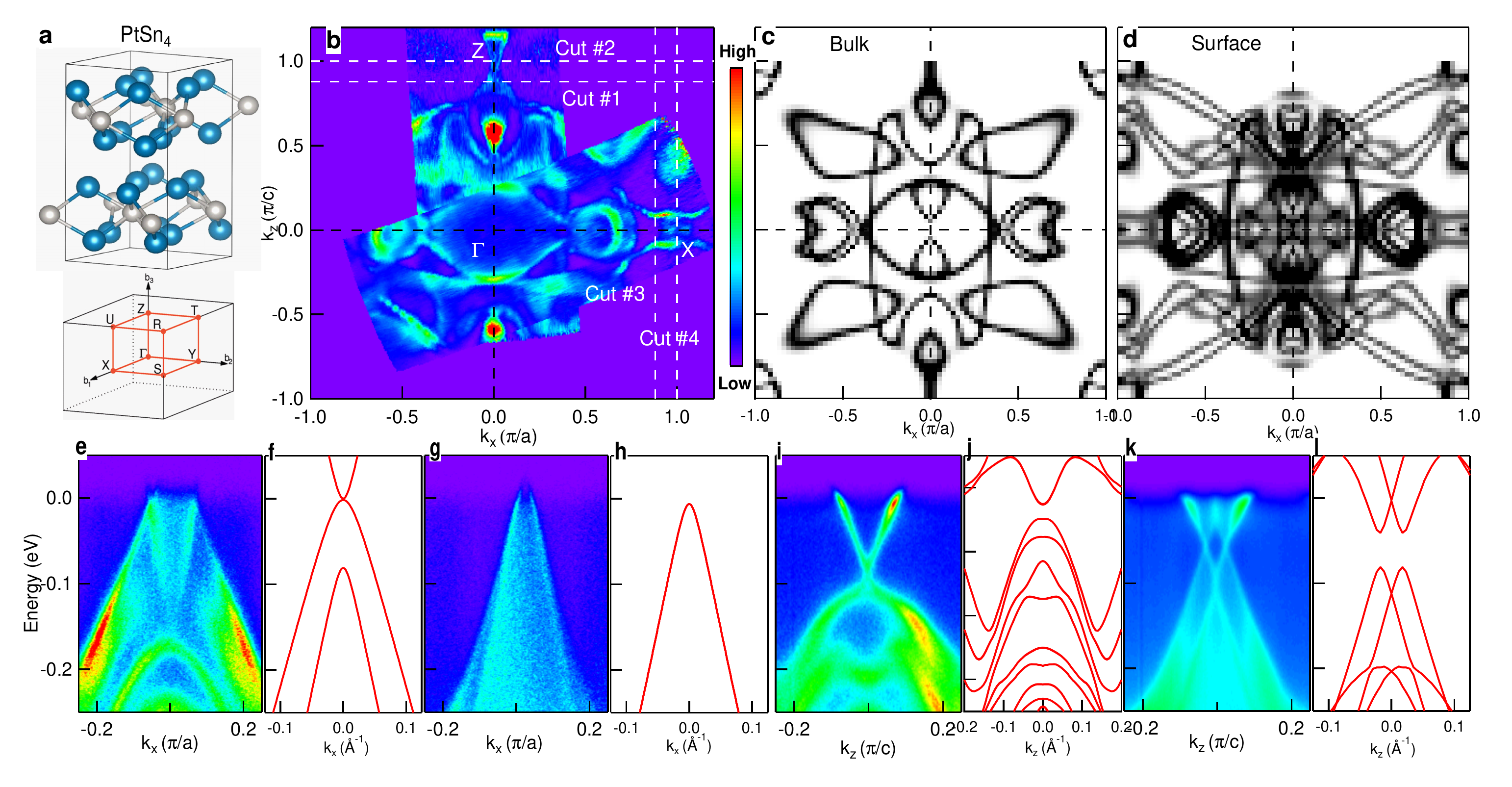}%
	\caption{(color online) Experimental and calculated structure of the Fermi surface and band dispersion of PtSn$_4$. 
	\textbf{a} Crystal structure (Pt: white spheres, Sn: blue spheres) and Brillouin zone of PtSn$_4$.
	\textbf{b} Fermi-surface plot of ARPES intensity integrated within 10 meV of the chemical potential along $\Gamma-Z$ and $\Gamma-X$.
	\textbf{c} DFT-calculated bulk FS slice at $k_{y} = 0.284 \pi/b$.
	\textbf{d} Calculated surface FS with a 24-layer slab.
	\textbf{e} Band dispersion at $k_z = 0.88 \pi/c$ along cut \#1 in \textbf{b}.
	\textbf{f} Calculated bulk-band dispersion at $k_z = 0.88 \pi/c \text{ and } k_{y} = 0.284 \pi/b$.
	\textbf{g} Band dispersion at $k_z = \pi/c$ along cut \#2 in \textbf{b}.
	\textbf{h} Calculated bulk-band dispersion at $k_z = \pi/c \text{ and } k_{y} = 0.284 \pi/b$.
	\textbf{i} Band dispersion at $k_x = 0.88 \pi/a$ along cut \#3 in \textbf{b}.
	\textbf{j} Calculated surface-band dispersion at $k_x = 0.88 \pi/a$ with 42-layer slab.
	\textbf{k} Band dispersion at $k_x = \pi/a$ along cut \#4 in \textbf{b}.
	\textbf{l} Calculated surface-band dispersion at $k_x = \pi/a$ with 42-layer slab.
	\label{fig:Fig1}}
\end{figure*}

\begin{figure}[bt]
	\includegraphics[scale=0.5]{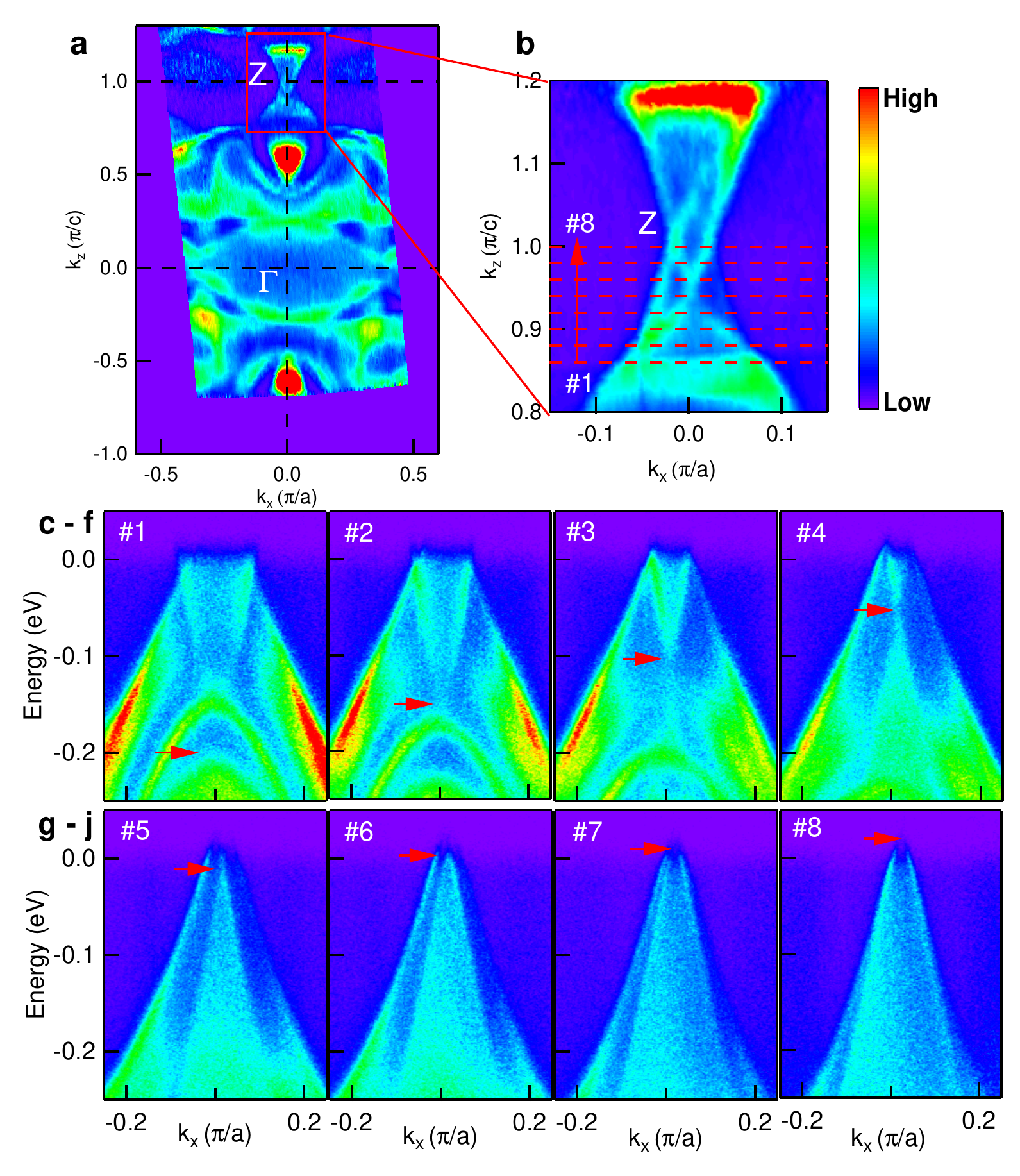}%
	\caption{(color online) Fermi surface and band dispersion in the proximity of the $Z$ point. 
	\textbf{a} Fermi surface plot of ARPES intensity integrated within 10 meV of the chemical potential along $\Gamma-Z$.
	\textbf{b} Zoom in image of the red box in Fig.\ref{fig:Fig2}\textbf{a}, red dashed lines mark the cut \#1 - \#8.
	\textbf{c - j} Band dispersion along cut \#1 - \#8. Cut \#8 is cutting through the $Z$ point. The red arrows mark the Dirac nodes.
	\label{fig:Fig2}}
\end{figure}

\begin{figure*}[bt]
	\includegraphics[scale=.5]{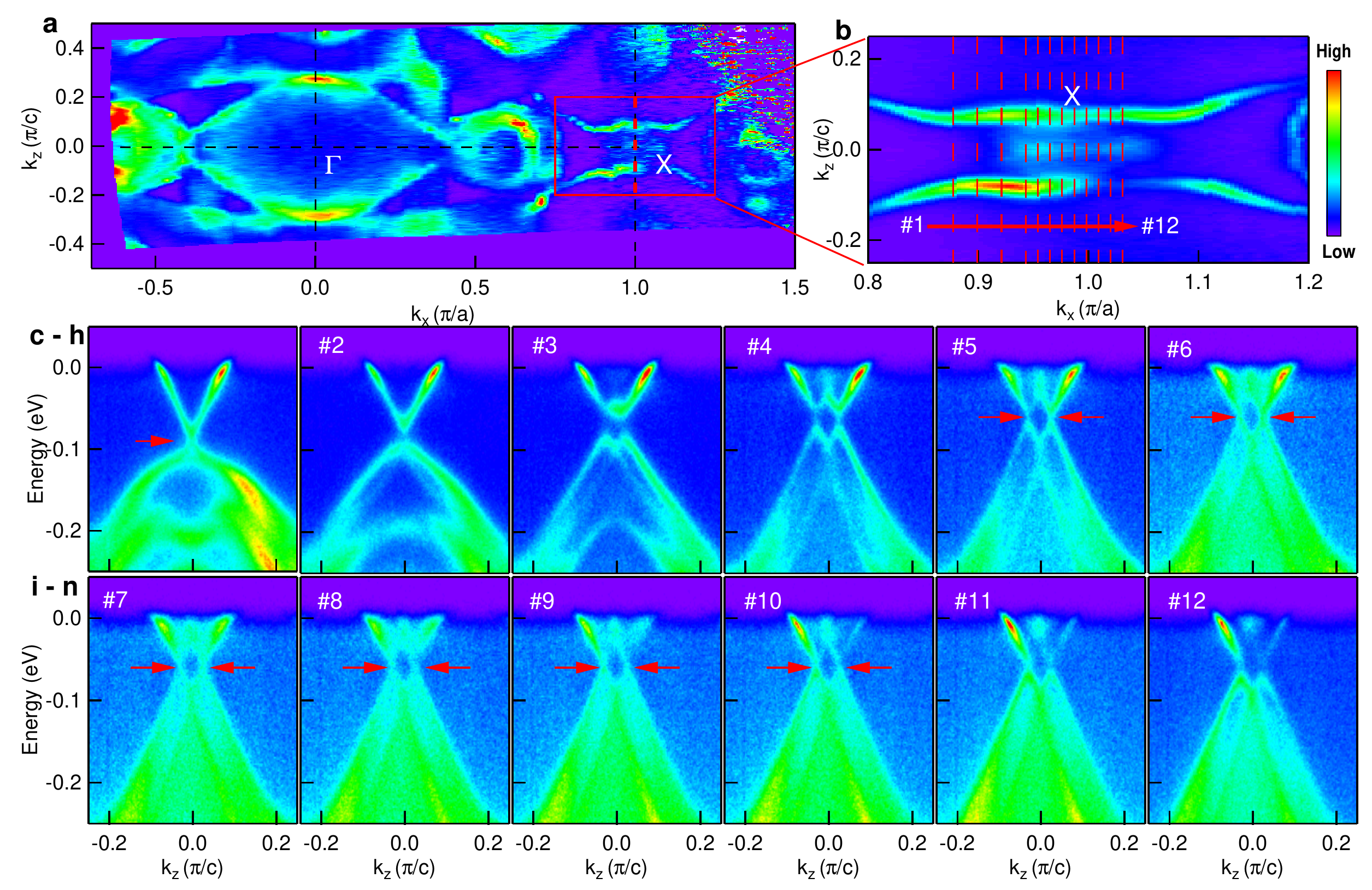}%
	\caption{(color online) Fermi Surface plot and band dispersion close to the $X$ point. 
	\textbf{a} Fermi surface plot of ARPES intensity integrated within 10 meV of the chemical potential along $\Gamma-X$.
	\textbf{b} Zoom in image of the red box in Fig.\ref{fig:Fig3}\textbf{a}, red dashed lines mark the cut \#1 - \#12.
	\textbf{c - n} Band dispersion along cut \#1 - \#12. Cut \#8 is cutting through the $X$ point. The red arrows mark the Dirac nodes.
	\label{fig:Fig3}}
\end{figure*}

\begin{figure}[tb]
	\includegraphics[scale=0.5]{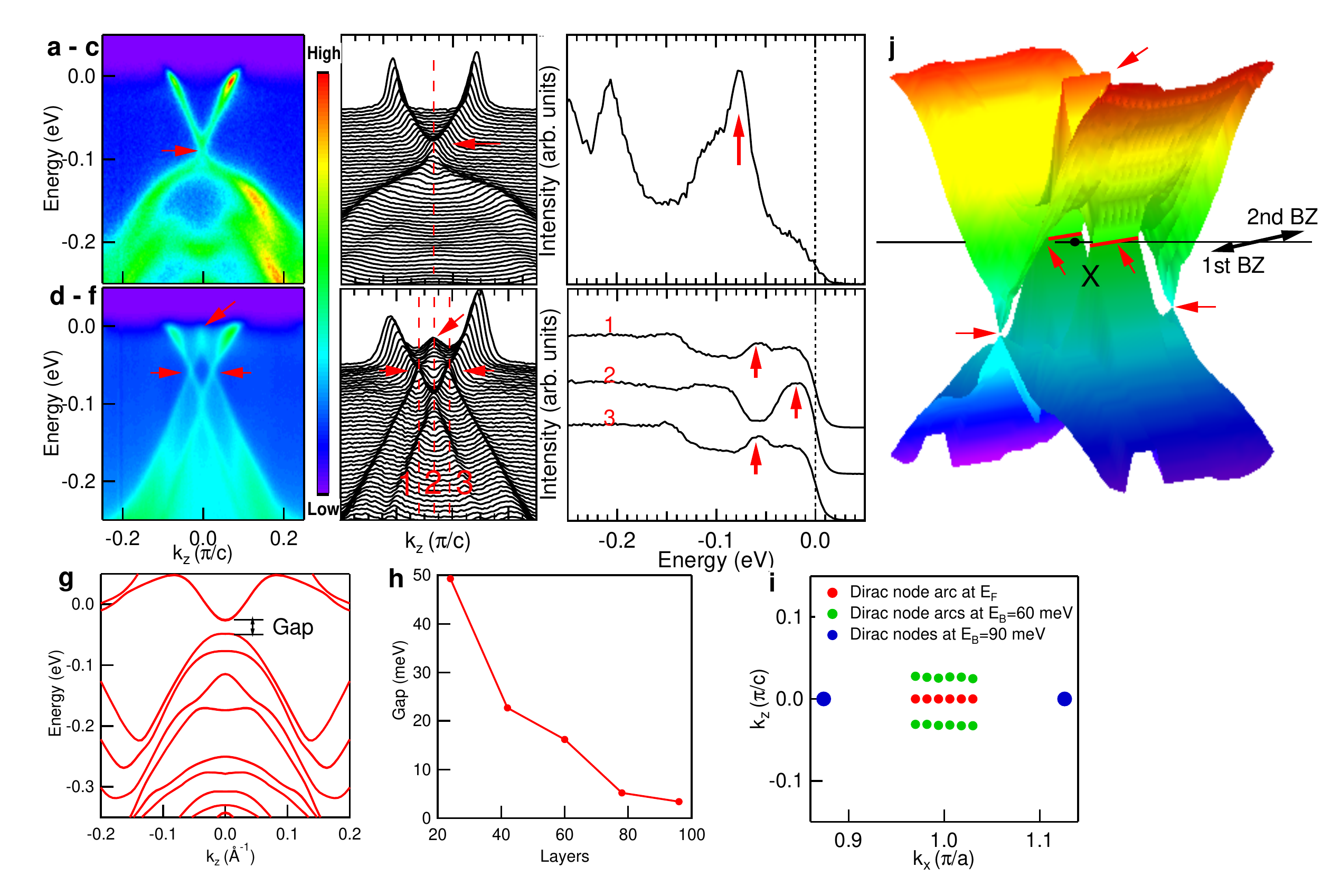}%
	\caption{(color online) Two types of gapless Dirac-like dispersion close to $X$ point. 
	\textbf{a} Band dispersion along cut \#1 in Fig.\ref{fig:Fig3}\textbf{b}.
	\textbf{b} Momentum Dispersion Curves (MDC) of \textbf{a}.
	\textbf{c} Energy Dispersion Curve (EDC) along the red dashed line in \textbf{b}.
	\textbf{d} Band dispersion along cut \#10 in Fig.\ref{fig:Fig3}\textbf{b}.
	\textbf{e} Momentum Dispersion Curves (MDC) of \textbf{d}.
	\textbf{f} Energy Dispersion Curves (EDC) along the red dashed lines in \textbf{e}.
	\textbf{g} Calculated surface band at $X$ with a 42-layer slab.
	\textbf{h} Energy-gap evolution with increasing number of layers in the slab.
	\textbf{i} Locations of the Dirac nodes extracted from the peak positions of the MDCs as marked by the red dashed lines in \textbf{a} and \textbf{d}. The blue dots denote the two single Dirac nodes at E$_B$ = 90 meV. The green dots denote the two Dirac node arcs at E$_B$ = 60 meV. The red dots denote the Dirac node arc at E$_F$.
	\textbf{j} Schematic of Double Dirac node arc structure. Red arrows mark the double single nodes and double node arcs. 
	\label{fig:Fig4}}
\end{figure}

The crystal structure, Fermi Surface and band dispersion along key directions in the Brillouin Zone (BZ) for PtSn$_4$ are shown in Fig.\ref{fig:Fig1}. Panel (\textbf{b}) shows the ARPES intensity integrated within 10 meV of the chemical potential. High intensity areas mark the contours of the FS sheets. The FS consists of at least one large electron pocket at the center of BZ surrounded by several other electron and hole FS sheets, consistent with the quantum oscillation result\cite{Mun12PRB}. Fig.\ref{fig:Fig1}\textbf{c} shows the calculated bulk FS, which matches the data well close to the center of the zone and $Z$ point in Fig.\ref{fig:Fig1}\textbf{b} and it is also consistent with the calculated FS using full potential linearized augmented plane wave (FLAPW) within local density approximation (LDA)\cite{Mun12PRB}. However, it does not predict the FS crossings close to the $X$ point, missing a set of nearly parallel FS sheets that are present in Fig.\ref{fig:Fig1}\textbf{b}. On the other hand, these experimental features are well reproduced by calculation of surface states using slab method; results of which are shown in Fig.\ref{fig:Fig1}\textbf{d}. Band dispersion along several cuts in proximity of $Z$ and $X$ points are shown in Fig.\ref{fig:Fig1}\textbf{e} - \textbf{l}. Close to $Z$ point (Figs.\ref{fig:Fig1}\textbf{e \& g}), the dispersion resembles a Dirac-like feature, but the intensity within band contour points to bulk origin and it is consistent with corresponding band calculations shown in Figs.\ref{fig:Fig1}\textbf{f, h}. Close to $X$ point (Figs.\ref{fig:Fig1}\textbf{i \& k}), the band dispersion is also Dirac-like, but very sharp, thus more likely to be due to surface states; moreover, it is consistent with the slab calculation shown in Figs.\ref{fig:Fig1}\textbf{j, l}. The data in Fig.\ref{fig:Fig1} demonstrates that the experimentally observed band structure has both bulk and surface components. The former dominate the Fermi surface close to the $Z$ point, and the latter is prominent close to the $X$ point.  The linear dispersion and gapless band crossings strongly suggest that both bulk and surface features at the edge of the Brillouin zone may have topological character possibly linked to ultrahigh magnetoresistance similar to Cd$_3$As$_2$\cite{Neupane14NatCom}, NbP\cite{Xu15SciAdv}, WTe$_2$\cite{Pan15arXiv, Wu15PRL}.

In Fig.\ref{fig:Fig2} we focus on the interesting features near the $Z$ point in more detail. Fig.\ref{fig:Fig2}\textbf{b} shows an enlarged image from the red box in Fig.\ref{fig:Fig2}\textbf{a}, where two triangular- shaped FS sheets are observed. Figs.\ref{fig:Fig2}\textbf{c - j} show the detailed evolution of band dispersions along cuts\#1 to \#8. A sharp linear dispersion starts to cross at binding energy of $\sim$200 meV in Fig.\ref{fig:Fig2}\textbf{c} and the Dirac point moves up in energy in Figs.\ref{fig:Fig2}\textbf{d} - \ref{fig:Fig2}\textbf{g} and finally reaches the Fermi level in Fig.\ref{fig:Fig2}\textbf{h}, as indicated by red arrows. Then, the Dirac point moves up above the Fermi level and becomes a sharp, shallow hole pocket in Figs.\ref{fig:Fig2}\textbf{g} - \ref{fig:Fig2}\textbf{j}. This movement of the Dirac nodes forms a line in the Energy-Momentum space in the proximity of $Z$. 

Whereas the behavior described above has previously been predicted by theory, the structure in the proximity of the $X$ point is far more interesting. We now examine the Fermi surface and band dispersion in small area in the part of the Brillouin zone that is marked by the red box in Fig.\ref{fig:Fig3}\textbf{a}. The Fermi surface in this region consists of a short arc along the the symmetry line and two longer, nearly parallel segments on either side of this arc. Detailed band dispersion along cuts\#1 to \#12 are shown in Figs.\ref{fig:Fig3}\textbf{c - n}. The data along cut \#1 shows Dirac-like dispersion, with the top and bottom bands merging at a single gapless point. The band is very sharp, consistent with its surface origin. As we move closer to the $X$ point, two things happen: a gap develops between top and bottom bands and both top and bottom bands split into two parts symmetric about the $k_z = 0$ line (Figs.\ref{fig:Fig3}\textbf{d - f}). Before reaching the $X$ point, the gap vanishes and there are two gapless Dirac-like features in close proximity of $X$ point. The inner bands of the two Dirac features merge along symmetry line and form the arc at the chemical potential visible in Fig.\ref{fig:Fig3}\textbf{b}. The two gapless, Dirac-like features extend along one direction in the proximity of the $X$ point between $k_x=0.95 \pi/a$ and $k_x=1.05 \pi/a$. Outside of this momentum range a gap develops separating the upper and lower portion of the band. This gives rise to two arcs of Dirac nodes located at binding energy of $\sim$ 60 meV, which we named Dirac node arcs.

We now proceed to demonstrate that the Dirac-like dispersion shown in Fig.\ref{fig:Fig3} is gapless by plotting the momentum-dispersion curves (MDCs) and energy-dispersion curves (EDCs). Fig.\ref{fig:Fig4}\textbf{c} shows the EDC extracted along the red dashed line in \ref{fig:Fig4}\textbf{b}. The red arrow marks the peak located at roughly 90 meV below the Fermi level (Fig.\ref{fig:Fig4}\textbf{a}), and demonstrates the absence of an energy gap in this single Dirac-like feature. In Fig.\ref{fig:Fig4}\textbf{d}, we show the double Dirac-like features along cut\#8(Fig.\ref{fig:Fig3}\textbf{j}). The EDCs shown in \ref{fig:Fig4}\textbf{f} are extracted along the red-dashed lines marked as 1, 2, 3 in Fig.\ref{fig:Fig4}\textbf{e}. The red arrows in Fig.\ref{fig:Fig4}\textbf{e} and Fig.\ref{fig:Fig4}\textbf{f} mark location of the peaks at binding energy of 60 meV, and show the gapless nature of these dispersions. The surface-state calculation using 42-layer slab shows that the conduction and valence bands are separated by roughly 23 meV in the single Dirac feature. However, further increase in the layer number would dramatically reduces the gap size, as shown in Fig.\ref{fig:Fig4}\textbf{h}. To better illustrate the Dirac node arc structure, we plot the location of the Dirac nodes in the momentum space in Fig.\ref{fig:Fig4}\textbf{i} by extracting the peak positions of MDCs at each node (i.e., at the binding energy of 90 meV in the single Dirac dispersion and 60 meV at the proximity of $X$ point in the double Dirac dispersion, as marked by the red-dashed lines in Fig.\ref{fig:Fig4}\textbf{a} and \textbf{d}). In Fig.\ref{fig:Fig4}\textbf{j}, the schematic of the double Dirac node arc structure is shown, with two Dirac dispersion extending along one-dimension in momentum space.

In conclusion, we use ultrahigh resolution tunable VUV laser-based ARPES to measure the Fermi surface and band dispersion of PtSn$_4$. The most significant result is the discovery of Dirac node arc structure in this material. Our results show that near the $X$ point, the single Dirac dispersion evolves into two gapped dispersions and, before reaching the $X$ point, the gaps close and two gapless Dirac-like feature emerge extending along one dimension in momentum space, forming Dirac node arc. These novel features differ from previously predicted Dirac line nodes that form closed loops in momentum space. We proposed that this novel topological nodal structure could be an ideal platform for studying the exotic properties of Dirac Fermions. 
Finally, we note that, most of the recently discovered ultrahigh magnetoresistive materials \cite{Liang15NatMat, Shekhar15NatPhys, Ali14Nat} seem to also possess Dirac or Weyl features in the band dispersions\cite{Neupane14NatCom, Xu15SciAdv, Pan15arXiv, Wu15PRL}. As we have demonstrated, this opens an avenue for discovering and identifying novel topological states and relativistic behavior based on rudimentary transport measurements.

\section*{Methods}
\textbf{Sample growth} Single crystals of PtSn$_4$ were grown out of a Sn-rich binary melt\cite{Canfield92PMPB}. The constituent elements, with an initial stoichiometry of Pt$_{0.04}$Sn$_{0.96}$, were placed in an alumina crucible and sealed in a quartz tube under a partial Ar pressure. After the quartz ampoule was heated up to 600~$^{\circ}\mathrm{C}$, the ampoule was cooled down to 350~$^{\circ}\mathrm{C}$ over 60 h\cite{Mun12PRB}. In order to decant the Sn readily at this temperature, a frit-disc crucible was used\cite{Canfield15arXiv}.

\textbf{Sample preparation and measurements} ARPES measurements were carried out using a laboratory-based system consisting of a Scienta R8000 electron analyzer and a a tunable VUV laser light source\cite{Jiang14RSI}. The data were acquired using a tunable VUV laser ARPES system, consisting of a Scienta R8000 electron analyzer, picosecond Ti:Sapphire oscillator and fourth harmaonic generator. All Data were collected with a constant photon energy of 6.7 eV. Momentum and energy resolution were set at $\sim$ 0.005 ~\AA$^{-1}$ and 2 meV. The size of the photon beam on the sample was $\sim$30 $\mu$m. Samples were cleaved \textit{in situ} at a base presure lower than $1 \times 10^{-10}$ Torr. Samples were cooled using a closed cycle He-refrigerator and the sample temperature was measured using a silicon-diode sensor mounted on the sample holder. The energy corresponding to the chemical potential was determinded from the Fermi edge of a polycrystalline Au reference in electrical contact with the sample. Samples were cleave at 40K and were kept at the cleaving temperature throughout the measurement. 

\textbf{Calculation method} Density functional theory\cite{Hohenberg64PR, Kohn65PR} (DFT) calculations have been done in VASP\cite{Kresse96PRB,Kresse96CMS} using PBE\cite{Perdew96PRL} exchange-correlation functional, plane-wave basis set with projected augmented waves\cite{Blochl94PRB} and spin-orbital coupling (SOC) effect included. For bulk band structure of PtSn$_4$, we use the conventional orthorhombic cell of 20 atoms with a Monkhorst-Pack\cite{Monkhorst76PRB} ($8 \times 6 \times 8$) $k$-point mesh. For surface band structure, we use slabs up to 96 atomic layers or 320 atoms with a ($8 \times 1 \times 8$) $k$-point mesh and at least a 12 \text{\AA} vacuum. The kinetic energy cutoff is 230 eV. The convergence with respect to $k$-point mesh was carefully checked, with total energy converged, e.g., well below 1 meV/atom. We use experimental lattice parameters\cite{Kunnen00JAC} of $a = 6.418\text{\AA}, b = 11.366 \text{\AA} \text{ and } c = 6.384 \text{\AA}$ with atoms fixed in their bulk positions.

\section*{Corresponding Author}
Correspondence to: Adam Kaminski, email: kaminski@ameslab.gov; Paul C. Canfield, email: canfield@ameslab.gov.

\section*{Acknowledgements}
This work was supported by the U.S. Department of Energy, Office of Science, Basic Energy Sciences, Materials Science and Engineering Division. Ames Laboratory is operated for the U.S. Department of Energy by Iowa State University under contract No. DE-AC02-07CH11358.

\section*{Author contributions}
P.C. C. initiated the work by insisting that Y. W. and A. K. design and carry out the experiment. Y. W., D. M. and L. H. acquired and analyzed ARPES data. L.-L. W., D.D. J. and Y. L. provided the density functional calculations. E. M. grew the samples under the supervision of S.L. B. and P.C. C.. Y. W. and A. K. wrote the draft of the manuscript. All authors discussed and commented on the manuscript.   

\section*{Additional information}
Supplementary information is available in the online version of the paper. Reprints and permissions information is available online at www.nature.com/reprints. 

\section*{Competing financial interests}
The authors declare no competing financial interests.

\end{document}